\documentclass[aps,pra,twocolumn,superscriptaddress,showpacs,10pt]{revtex4-1}
\pdfoutput=1
\usepackage{amsmath}
\usepackage{graphicx}
\usepackage{epstopdf}
\usepackage[normalem]{ulem}

\begin{document}

\title{Coherent control of a single trapped Rydberg ion}

\author{Gerard Higgins}
\email[]{gerard.higgins@fysik.su.se}
\affiliation{Department of Physics, Stockholm University, SE-106 91 Stockholm, Sweden}
\affiliation{Institut f\"ur Experimentalphysik, Universit\"at Innsbruck, AT-6020 Innsbruck, Austria}
\author{Fabian Pokorny}
\affiliation{Department of Physics, Stockholm University, SE-106 91 Stockholm, Sweden}
\author{Chi Zhang}
\affiliation{Department of Physics, Stockholm University, SE-106 91 Stockholm, Sweden}
\author{Quentin Bodart}
\affiliation{Department of Physics, Stockholm University, SE-106 91 Stockholm, Sweden}
\author{Markus Hennrich}
\affiliation{Department of Physics, Stockholm University, SE-106 91 Stockholm, Sweden}
\email[]{markus.hennrich@fysik.su.se}

\date{\today}


\maketitle

\textbf{Trapped Rydberg ions are a promising novel approach to quantum computing and simulations \cite{Mueller2008, Feldker2015, Higgins2017}. They are envisaged to combine the exquisite control of trapped ion qubits \cite{Schindler2013b} with the fast two-qubit Rydberg gates already demonstrated in neutral atom experiments \cite{Isenhower2010, Browaeys2016, Saffman2016}. Coherent Rydberg excitation is a key requirement for these gates. Here, we carry out the first coherent Rydberg excitation of an ion and perform a single-qubit Rydberg gate, thus demonstrating basic elements of a trapped Rydberg ion quantum computer.
}

Systems of trapped ion qubits have set numerous benchmarks for single-qubit preparation, manipulation, and readout \cite{Harty2014}. They can perform low error entanglement operations \cite{Ballance2016, Gaebler2016} with up to 14 ion qubits \cite{Monz2011}. Still, a major limitation towards realizing a large-scale trapped ion quantum computer or simulator is the scalability of entangling quantum logic gates \cite{Monroe2013}.

Arrays of neutral atoms in dipole traps offer another promising approach to quantum computation and simulation. Here, qubits are stored in electronically low-lying states and multi-qubit gates may be realized by exciting atoms to Rydberg states \cite{Jaksch2000, Saffman2010, Browaeys2016, Saffman2016}. Rydberg states are exotic states of matter in which the valence electron is excited to high principal quantum numbers.
They can have extremely high dipole moments and may interact strongly with each other, which has allowed entanglement generation \cite{Wilk2010, Jau2016} and fast two-qubit Rydberg gates \cite{Isenhower2010} in neutral atom systems.

A system of trapped Rydberg ions may combine the advantages of both technologies. Electronically low-lying states may be used as qubit states and fast multi-qubit gates are envisaged by coherently exciting ions to Rydberg states and employing dipolar interactions between them \cite{Mueller2008, Li2014}. Multi-qubit gates commonly used in trapped ion systems suffer scalability restrictions due to spectral crowding of motional modes \cite{Monroe2013}. This issue does not affect multi-qubit Rydberg gates thus a trapped Rydberg ion quantum computer offers an alternate approach to a scalable system.

An unanswered question was whether trapped ions can be excited to Rydberg states in a coherent fashion as is required for multi-qubit Rydberg gates. In this work we perform coherent Rydberg excitation of a single trapped ion by stimulated Raman adiabatic passage (STIRAP) with $(91\pm 3)\%$ transfer efficiency. We combine coherent Rydberg excitation by STIRAP with qubit manipulation to demonstrate a single-qubit Rydberg phase gate. We perform process tomography of the gate with a $\pi$-phase shift and measure $(78\substack{+4 \\ -8})\%$ fidelity.
This indicates our system is capable of implementing a two-qubit Rydberg gate (as proposed in \cite{Moller2007}) using dipolar interactions between microwave-dressed Rydberg states \cite{Mueller2008, Li2014}, thus a trapped Rydberg ion quantum computer may be feasible.

\begin{figure}
	\includegraphics[width=0.8\columnwidth]{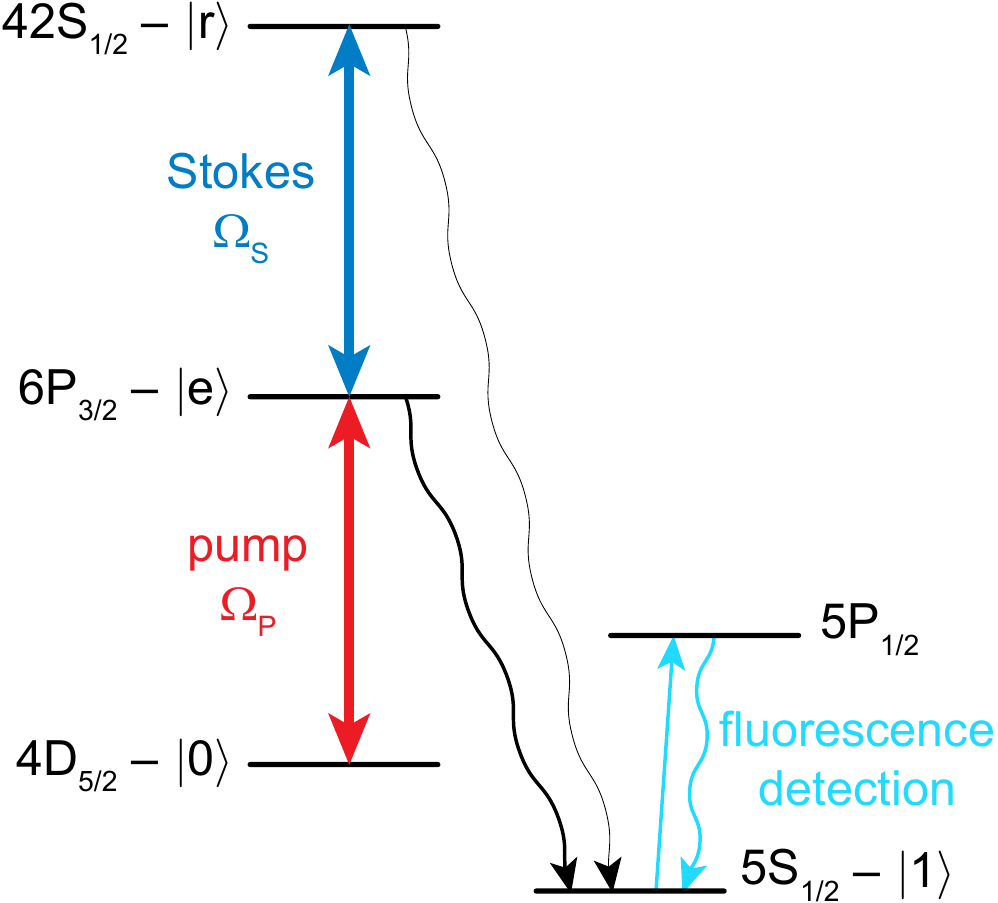}
	\caption{
		\textbf{Three atomic levels of \textsuperscript{88}Sr\textsuperscript{+} are coupled by two UV lasers.}
		The qubit state $|0\rangle$ is coupled to $|e\rangle$ by the pump laser, $|e\rangle$ is coupled to the Rydberg state $|r\rangle$ using the Stokes laser.
		Population in $|e\rangle$ or $|r\rangle$ decays mostly to $5S_{1/2}$; detection of scattered fluorescence light heralds excitation from $|0\rangle$ and decay to $5S_{1/2}$.
		\label{Fig1-EIT}}
\end{figure}

In our experiment we study a single \textsuperscript{88}Sr\textsuperscript{+} ion confined in a linear Paul trap. Three atomic levels in a ladder configuration are coupled using two UV lasers (Fig.~\ref{Fig1-EIT}). The qubit state $|0\rangle$ is coupled to the excited state $|e\rangle$ by the pump laser at 243\,nm with Rabi frequency $\Omega_{P}$. $|e\rangle$ is coupled in turn to the Rydberg state $|r\rangle$ ($42S_{1/2}, m_J=-1/2$) using the Stokes laser at 307\,nm with Rabi frequency $\Omega_{S}$. The experimental setup is described in detail in the Methods section and in \cite{Higgins2017}.

\begin{figure*}
	\includegraphics[width=0.9\textwidth]{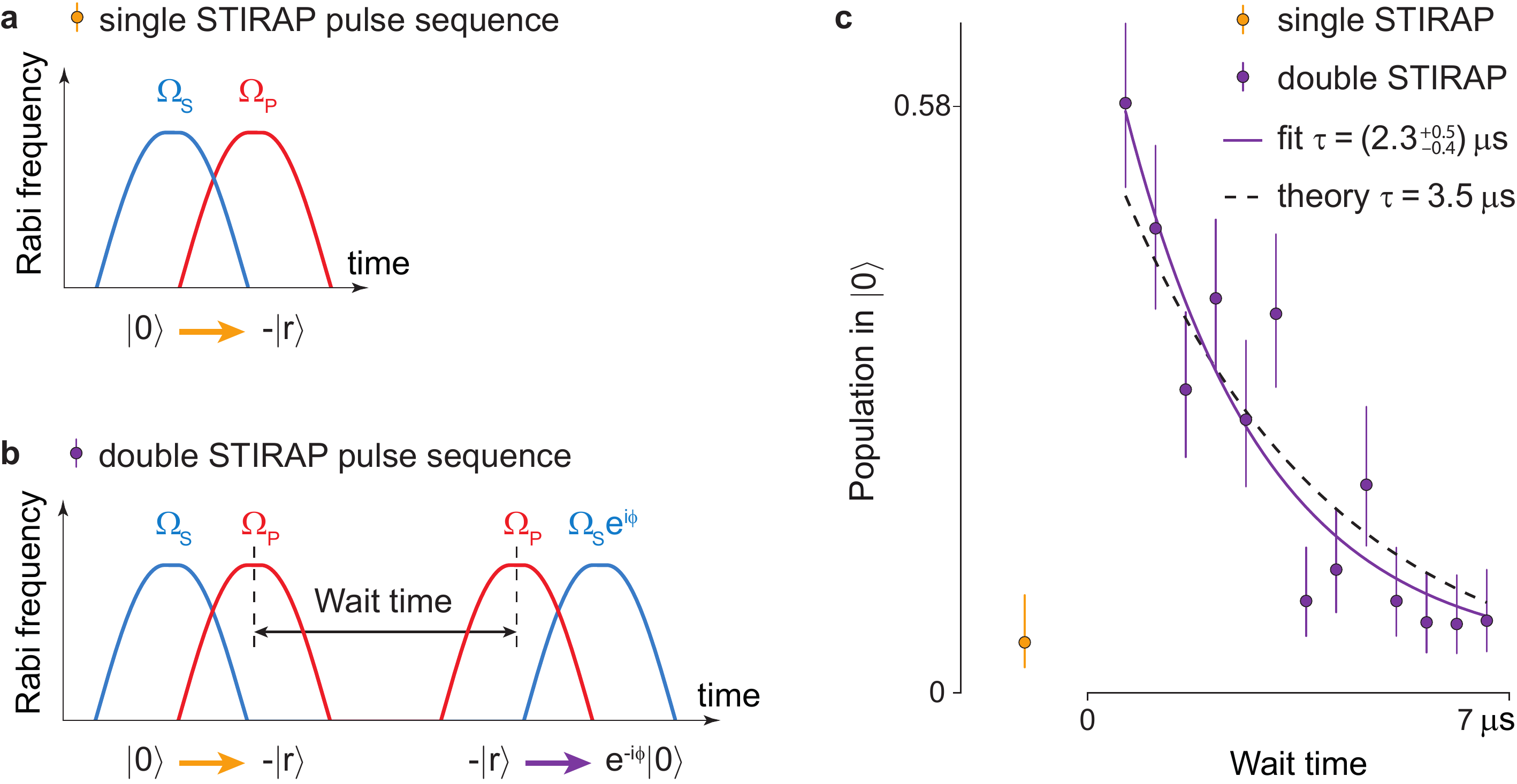}%
	\caption{
		\textbf{Coherent Rydberg excitation by STIRAP shown by comparing application of the single and the double STIRAP pulse sequences.}
		The single STIRAP pulse sequence, \textbf{a,} transfers nearly all the population from $|0\rangle \rightarrow 5S_{1/2}$ via $|r\rangle$ or $|e\rangle$, shown in \textbf{c}.
		After the double STIRAP pulse sequence, \textbf{b,} most of the population may be returned to $|0\rangle$, as in \textbf{c,} which shows population is transferred successfully to the Rydberg state $|r\rangle$.
		By varying the wait time between the two sets of STIRAP pulses the Rydberg state lifetime is measured.
		Error bars indicate projection noise (68\% CI).
		The evolution of the dark state $|\Phi_{dark}\rangle$ is described beneath each pulse sequence.
		\label{Fig2-Stirap}}
\end{figure*}

We can use the two-photon coupling for coherent control of the Rydberg excitation.
At two-photon resonance ($|0\rangle$ to $|r\rangle$) the coupling Hamiltonian has a ``dark" eigenstate $|\Phi_{dark}\rangle \sim \Omega_{S} e^{i\phi} |0\rangle - \Omega_{P} |r\rangle$ (Methods), which is named so because it does not contain any component of the lossy state $|e\rangle$ and thus it does not scatter light in timescales much less than $\tau_{42S}$ the lifetime of the Rydberg state $42S_{1/2}$ ($|r\rangle$). 
The character of the dark state depends on the ratio $\tfrac{\Omega_{S}}{\Omega_{P}}$, thus by adiabatically varying $\Omega_{S}$ and $\Omega_{P}$ according to the pulse sequence in Fig.~\ref{Fig2-Stirap}a, population initially in $|0\rangle$ follows the evolution of the dark state and is transferred to $-|r\rangle$, without populating $|e\rangle$.
This process is called stimulated Raman adiabatic passage (STIRAP) \cite{Vitanov2017}.

Laser pulses are shaped by driving acousto-optic modulators using arbitrary waveform generators such that the Rabi frequencies rise and fall in a sinusoidal fashion, as in Fig.~\ref{Fig2-Stirap}, over $t_{\mathrm{rise}}=200\,\mathrm{ns}$.
$t_{\mathrm{rise}}$ is chosen to be long enough for the state vector to evolve adiabatically, while kept short to reduce losses from Rydberg state decay and from decoherence due to finite laser linewidths.
The peak Rabi frequencies of the two lasers are matched (Methods) and made as high as is experimentally attainable to maximize the STIRAP efficiency $\{\Omega_{P}, \Omega_{S}\} \sim 2 \pi \times 47\,\mathrm{MHz}$.
Both UV laser linewidths are estimated to be $\sim 2 \pi \times 100\,\mathrm{kHz}$ and are sufficiently low ($\ll \Omega_{P}, \Omega_{S}, \tfrac{1}{t_{\mathrm{rise}}}$) to allow adiabatic following.

\begin{figure*}
	\includegraphics[width=0.9\textwidth]{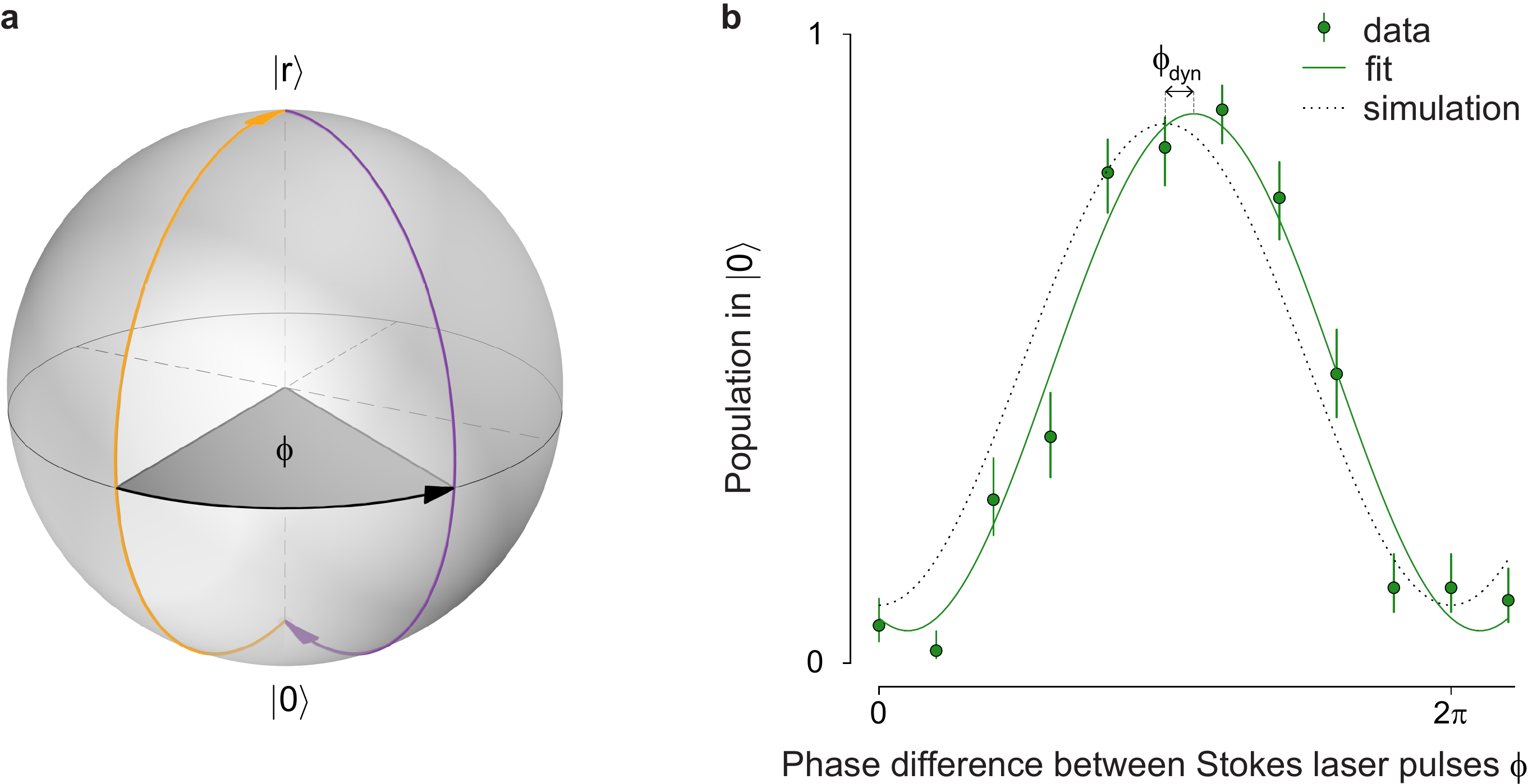}
	\caption{
		\textbf{Accumulation of a geometric phase during the double STIRAP sequence.}
		\textbf{a,} We control the path transversed by the dark state on the Bloch sphere spanned by $|0\rangle$ and $|r\rangle$ by varying $\phi$ in Fig.~\ref{Fig2-Stirap}b.
		The area enclosed by the path results in the accumulation of a geometric phase $|0\rangle \rightarrow e^{-i\phi}|0\rangle$.
		\textbf{b,} The geometric phase is measured in a Ramsey experiment.
		The double STIRAP sequence is nested between two Ramsey pulses on the 674\,nm transition between the qubit states $|0\rangle\leftrightarrow|1\rangle$.
		As $\phi$ is varied the final population in $|0\rangle$ oscillates.
		Error bars indicate projection noise (68\% CI).
		\label{Fig3-Stirap-phase}}
\end{figure*}
Detecting population transfer by STIRAP in our system relies upon discriminating population in the state $5S_{1/2}$ from population in the initial state $|0\rangle$ by detecting fluorescence on the $5S_{1/2}\leftrightarrow5P_{1/2}$ transition (Fig.~\ref{Fig1-EIT}).
After applying the single STIRAP pulse sequence (Fig.~\ref{Fig2-Stirap}a) nearly all the population is transferred out of $|0\rangle$ into $5S_{1/2}$ (Fig.~\ref{Fig2-Stirap}c), since population decays from both $|r\rangle$ and $|e\rangle$ overwhelmingly to $5S_{1/2}$.
Here we cannot distinguish successful STIRAP from simple optical pumping via $|e\rangle$.
We therefore apply a double STIRAP sequence which concludes with 58\% of the population returned to  $|0\rangle$, thus demonstrating successful excitation and de-excitation by STIRAP.
The return of population is not perfect because the state vector does not perfectly follow the dark state throughout the pulse sequence, due to Rydberg state decay, finite laser linewidths and short $t_{\mathrm{rise}}$.

During the wait time in the double STIRAP sequence population may decay from $|r\rangle$ and be removed from the three-level system.
By measuring the population returned to $|0\rangle$ as the wait time is varied the lifetime of $|r\rangle$ is determined $\tau_{42S}=(2.3\substack{+0.5 \\ -0.4})\,\mathrm{\mu s}$ (Fig.~\ref{Fig2-Stirap}c).
This is the first lifetime measurement of a trapped Rydberg ion.
Comparing this with the theoretically-predicted lifetime of $42S_{1/2}$ in free space at 300\,K $3.5\,\mathrm{\mu s}$ (see Methods section) suggests the Rydberg state lifetime is not significantly shortened by confinement in the Paul trap.
This is important for trapped Rydberg ions to be a viable quantum technology, since Rydberg state lifetimes place fundamental limits on gate fidelities \cite{Saffman2010} and resonance linewidths.

When the wait time in the double STIRAP sequence (Fig.~\ref{Fig2-Stirap}b) was set to zero, $(83\substack{+5 \\ -6})\%$ of the population was returned to $|0\rangle$, which indicates a single STIRAP efficiency of $\sqrt{(83\substack{+5 \\ -6})\%} = (91\pm3)\%$.
This marks a significant improvement on the highest STIRAP efficiency observed with neutral Rydberg atoms (60\%) \cite{Cubel2005, Deiglmayr2006, Sparkes2016}.

We use the double STIRAP sequence to introduce a geometric phase following the protocol recently demonstrated with a solid-state qubit \cite{Yale2016}. During the double STIRAP sequence, the dark state moves on the surface of the Bloch sphere spanned by $|0\rangle$ and $|r\rangle$ from the $|0\rangle$ pole to the $|r\rangle$ pole then back to $|0\rangle$ (Fig.~\ref{Fig3-Stirap-phase}a).
When the dark state reaches the $|r\rangle$ pole the phase of the Stokes laser is shifted by $\phi$ using an acousto-optic modulator, and the dark state returns to the $|0\rangle$ pole along a different Bloch sphere longitude.
This `tangerine slice' trajectory with wedge angle $\phi$ circumscribes a solid angle $2\phi$ and gives rise to an accumulated geometric phase $-\phi$.

This geometric phase is detected by using the other qubit state $|1\rangle$ ($5S_{1/2}, m_{J}=-\frac{1}{2}$) as a phase reference in a Ramsey experiment.
The double STIRAP sequence is nested between two Ramsey pulses on the $|1\rangle \leftrightarrow |0\rangle$ transition at  674\,nm.
A complete oscillation in the $|0\rangle$ population is observed as $\phi$ is varied, Fig.~\ref{Fig3-Stirap-phase}b, which shows an arbitrary geometric phase may be acquired.
The imperfect STIRAP efficiency and additional decoherence from finite UV laser linewidths cause the contrast of the oscillation, defined as the maximal value minus the minimal value, to be less than unity $\mathcal{C}= (82 \pm 4) \%$.
Decay of population outside the \{$|0\rangle,|1\rangle$\} manifold to $5S_{1/2}, m_{J}=+\frac{1}{2}$ causes the center of the oscillation to be lower than 0.5.
This experiment is simulated using the Lindblad master equation with experimentally determined parameters.
Excellent agreement is observed between the simulation and the experimental results.
The $(18\pm4)^{\circ}$ dynamic phase offset $\phi_{dyn}$ in the experimental data may be accounted for by small detunings ($\sim 2 \pi \times 100\,\mathrm{kHz}$) of the UV lasers from resonance and the light shift from the Stokes laser acting on $|1\rangle$.
The simulated curve displays a lower contrast than the experimental results, which may be accounted for by overestimation of the UV laser linewidths.
The simulation predicts a STIRAP efficiency of 90\%, limited by Rydberg state decay, UV laser linewidths and non-adiabaticity in the state evolution during the short rise time.
Thus the efficiency may be improved by increasing the UV laser light intensity at the ion to allow faster adiabatic passage, improving the UV laser stabilization and exciting higher Rydberg states with longer lifetimes \cite{Gallagher2005}.

\begin{figure}
	\includegraphics[width=\columnwidth]{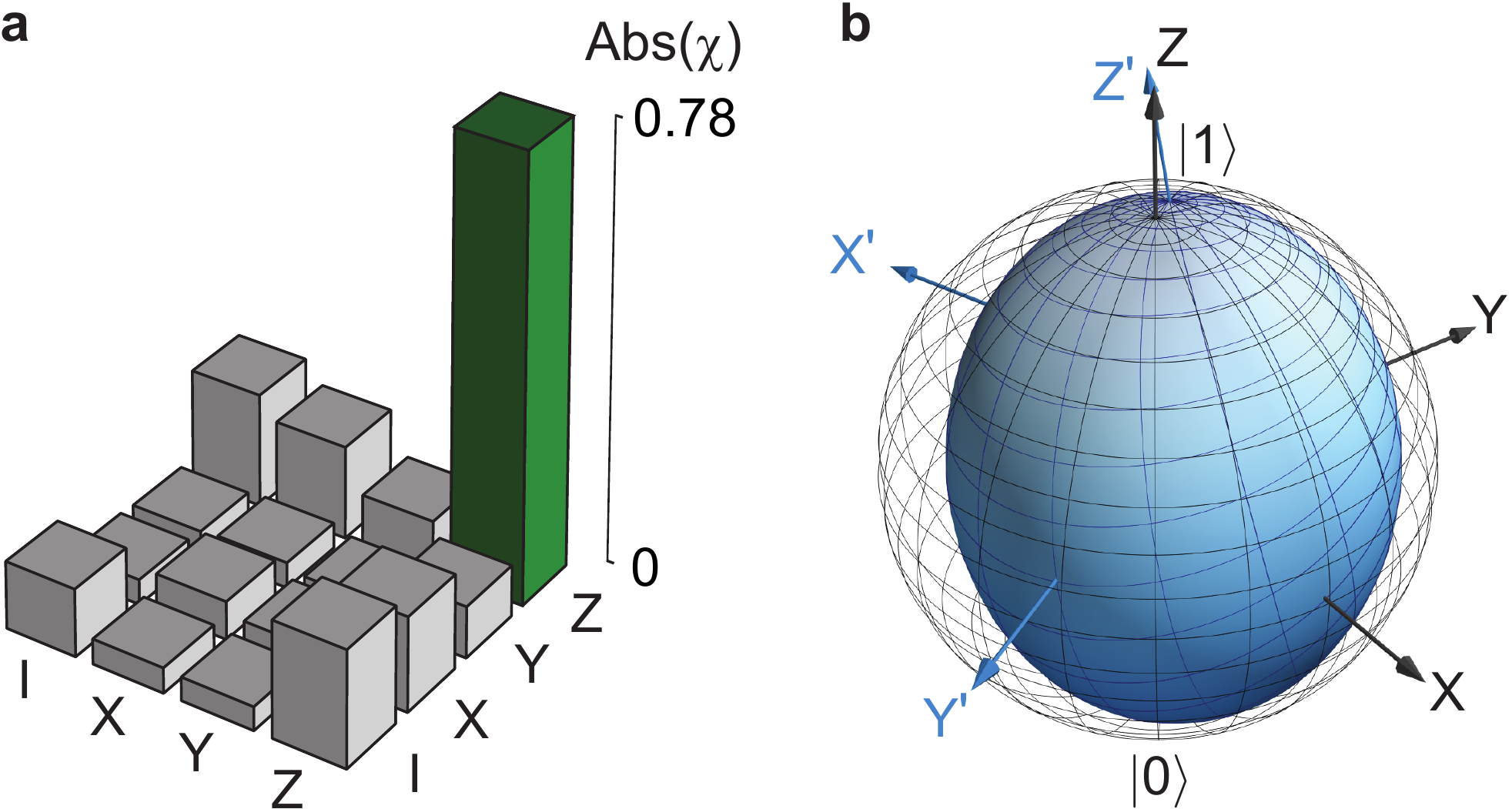}
	\caption{
		\textbf{Process tomography results of the double STIRAP sequence (Fig.~\ref{Fig2-Stirap}b) with $\phi=\pi$.}
		\textbf{a,} The absolute value of the process matrix. The process fidelity $(78^{+4}_{-8})$\,\% is the height of the green bar, which is identical to a $\sigma_{z}$-rotation.
		\textbf{b,} Reconstructed Bloch sphere after the sequence. In the ideal process the sphere would be rotated about the z-axis by $\pi$.
		\label{Fig4-process_tomo}}
\end{figure}

The double STIRAP sequence that introduces a geometric phase implements a single-qubit phase gate in the qubit basis. We characterize this operation for $\phi=\pi$, (double STIRAP $|0\rangle \rightarrow -|r\rangle \rightarrow -|0\rangle$) by performing quantum process tomography using a maximum likelihood estimation \cite{Chuang1997, James2001}.
The measured process fidelity is $(78\substack{+4 \\ -8})$\,\%, the errors are estimated using Monte Carlo simulations.
The reconstructed process matrix $\chi$ and the post-process Bloch sphere are plotted in Fig.~\ref{Fig4-process_tomo}.
A perfect operation with process fidelity unity would result in a $\sigma_{z}$-rotation and the post-process Bloch sphere would be rotated about the z-axis by $\pi$.
Imperfections in the process cause a shrinking of the $|0\rangle$ pole of the Bloch sphere due to the imperfect STIRAP efficiency and a rotation about the z-axis $\neq \pi$ due to a dynamical phase shift. The post-process Bloch sphere is not symmetric about the z-axis because of imperfect Ramsey pulses.

We experimentally realized coherent Rydberg excitation of a single trapped ion and implemented a single-qubit geometric phase gate in Rydberg excitation and deexcitation.
These are basic elements of a trapped Rydberg ion quantum computer. In particular, the single-qubit Rydberg phase gate demonstrated here may be extended to a two-qubit controlled phase gate \cite{Moller2007} based on Rydberg interaction \cite{Mueller2008, Li2014}.
Such a gate offers an alternate approach towards a scalable system and may open up a new paradigm for quantum computation.

\section*{Acknowledgments}
The research leading to these results has received funding from the European Research Council under the European Union's Seventh Framework Programme (FP/2007-2013) / ERC Grant Agreement n. 279508.

\setcounter{equation}{0}
\setcounter{figure}{0}
\setcounter{table}{0}
\makeatletter
\renewcommand{\theequation}{M\arabic{equation}}
\renewcommand{\thefigure}{M\arabic{figure}}

\section*{Methods \label{sect:methods}}
Three atomic levels are coupled by two UV lasers.
The stretched states $4D_{5/2}, m_{J}=-\frac{5}{2}$ ($|0\rangle$), $6P_{3/2}, m_{J}=-\frac{3}{2}$ ($|e\rangle$) and $42S_{1/2}, m_{J}=-\frac{1}{2}$ ($|r\rangle$) are employed to achieve an effective three-level system through the electric dipole selection rules.
Transitions between stretched states have the largest Clebsch-Gordan coefficients and thus allow the highest Rabi frequencies to be attained. 
The two UV lasers counterpropagate along the trap axis thus Doppler effects are minimized. Their polarizations allow excitation of only $\mathrm{\sigma^{+}}$ transitions. The degeneracy of Zeeman sublevels is lifted by a magnetic field oriented along the trap axis with field strength $0.36\,\mathrm{mT}$. $5S_{1/2}, m_{J}=-\frac{1}{2}$ is employed as the other qubit state, which is labeled $|1\rangle$. Population decays from $|r\rangle$ and $|e\rangle$ to both Zeeman sublevels of $5S_{1/2}$.

The ion is trapped with axial trapping frequency $2 \pi \times 872\,\mathrm{kHz}$ and radial trapping frequencies $2 \pi \{1.645, 1.707 \} \,\mathrm{MHz}$.
The trap drive is $2 \pi \times 18.15\,\mathrm{MHz}$.

In the experiments reported in the main text, the ion is Doppler cooled in the axial and radial directions.
In addition radial sideband cooling is employed to minimize dephasing which results from coupling between electronic and motional dynamics in the Rydberg state \cite{Mueller2008, Higgins2017}.

Experimental sequences are typically repeated 50 times for each measurement point.
Usually we carry out hundreds of experiments with the same ion before it is lost by double-ionization.

\renewcommand{\citenumfont}[1]{M#1}
To maximize the STIRAP transfer efficiency, the Rabi frequencies $\Omega_{P}$, $\Omega_{S}$ are matched and the two-photon resonance condition ($|0\rangle$ to $|r\rangle$) is met as follows:
The Lorentzian absorption profile of $|0\rangle \leftrightarrow |e\rangle$ is observed by scanning the frequency of the pump laser across the resonance in the absence of the Stokes laser (Supplementary material).
A weak pump laser is used $\Omega_{P}\ll\Gamma_{e}$, $\Gamma_{e}$ is the $|e\rangle$ linewidth.
We measure $\Gamma_{e} = 2 \pi \times (4.5 \pm 0.5) \,\mathrm{MHz}$, which is consistent with theoretical values for the natural decay rate of $|e\rangle$ \cite{Biemont2000, Safronova2010}.
Fitting the absorption profile allows $\Omega_{P}$ and the detuning of the pump laser $\Delta_{P}$ to be determined.
When the strong Stokes laser is turned on and the frequency scan of the pump laser repeated the absorption profile shows an Autler-Townes doublet.
Fitting this absorption profile allows $\Omega_{S}$ and the detuning of the Stokes laser $\Delta_{S}$ to be determined.
The $|0\rangle \leftrightarrow |e\rangle$ absorption profile and absorption profiles showing Autler-Townes splitting are included in the Supplementary material.
The two-photon resonance condition required for STIRAP is met by setting the detunings of both UV lasers to zero $\Delta_{P}=-\Delta_{S}=0$.
In this calibration procedure typically only Doppler cooling is employed.

Population transfer by STIRAP occurs provided the state vector adiabatically follows the dark state $|\Phi_{dark}\rangle$ as the dark state evolves.
The dark state is found as follows:
The three levels $\{|0\rangle, |e\rangle, |r\rangle\}$ are coupled by the two UV lasers according to the coupling Hamiltonian $\mathsf{H}$ in the rotating wave approximation
\begin{equation*}
\mathsf{H}=\frac{\hbar}{2}
\begin{pmatrix}
    0          & \Omega_{P}            & 0                       \\
    \Omega_{P} & 2\Delta_{P}           & \Omega_{S} e^{i\phi}    \\
    0          & \Omega_{S} e^{-i\phi} & 2\Delta_{P}+2\Delta_{S}
\end{pmatrix}
\end{equation*}
in the basis $\{|0\rangle, |e\rangle, |r\rangle\}$. At two-photon resonance ($\Delta_{P}=-\Delta_{S}$) one of the eigenstates of $\mathsf{H}$ is the dark state $|\Phi_{dark}\rangle \sim \Omega_{S} e^{i\phi} |0\rangle - \Omega_{P} |r\rangle$.

\renewcommand{\citenumfont}[1]{#1}
The theoretical lifetime of the $42S_{1/2}$ Rydberg state ($|r\rangle$) at T=300\,K in Fig.~\ref{Fig2-Stirap}c is found by extending the calculations for the lifetime at T=0\,K in \cite{Higgins2017} to include transitions driven by blackbody radiation \renewcommand{\citenumfont}[1]{M#1}\cite{Gallagher1979}.

To determine the best fit parameters and their uncertainties for the data in Figs. \ref{Fig2-Stirap}c and \ref{Fig3-Stirap-phase}b the posterior distribution functions are obtained by multiplying the likelihood functions with flat priors, then they are sampled using an ensemble Markov chain Monte Carlo sampler \cite{Foreman-Mackey2013}.

In the simulation in Fig.~\ref{Fig3-Stirap-phase}b the Lindblad master equation for the five-level system \{$|0\rangle, |e\rangle, |r\rangle, |5S_{1/2}, m_{J}=+1/2\rangle, |5S_{1/2}, m_{J}=-1/2\rangle \equiv |1\rangle$\}
is solved numerically with the open source Python framework QuTiP \cite{Johansson2013}. The Zeeman sublevels of $5S_{1/2}$ are treated separately in the simulation since the Ramsey pulses couple $|0\rangle$ with only one of the sublevels $5S_{1/2}, m_{J}=-1/2$ ($|1\rangle$). When the values of the UV laser linewidths used in the simulation are decreased to $2 \pi \times 64\,\mathrm{kHz}$ the contrast of the simulated results matches the experimentally-observed contrast and the simulation returns STIRAP efficiency 91\%. 

\renewcommand{\bibnumfmt}[1]{[M#1]}

\setcounter{equation}{0}
\setcounter{figure}{0}
\setcounter{table}{0}
\makeatletter
\renewcommand{\theequation}{S\arabic{equation}}
\renewcommand{\thefigure}{S\arabic{figure}}

\section*{Supplementary Material \label{sect:Supplementary}}
Coherent Rydberg excitation by STIRAP requires precise control of intensity and frequency of the excitation lasers. We measure Rabi frequencies and detunings by characterizing the Autler-Townes splitting (EIT) in two-photon excitation. Details are described in the following.

A single Doppler-cooled ion is initialized in $|0\rangle$ and excited to $|e\rangle$ or $|r\rangle$ or a superposition of both states using the UV lasers (see Fig.~1 in the main text).
From $|e\rangle$ or $|r\rangle$ more than 93\% of the population decays to the $5S_{1/2}$ ground state by multi-step decay processes in \textless$\sim20\,\mu s$ \renewcommand{\citenumfont}[1]{S#1}\cite{Zhang2016}, \renewcommand{\citenumfont}[1]{M#1}\cite{Safronova2010}, \renewcommand{\citenumfont}[1]{#1}\cite{Higgins2017}.
Fluorescence detection lasers which drive the $5S_{1/2}\leftrightarrow5P_{1/2}$ transition are then switched on; detection of scattered fluorescence light heralds excitation from $|0\rangle$ followed by decay to $5S_{1/2}$.

In this section, the pump laser Rabi frequency is much lower than the Stokes laser Rabi frequency and the linewidth of $|e\rangle$, $\Omega_{P}\ll\Omega_{S},\Gamma_{e}$. In this regime it is conventional to call the pump and Stokes lasers the probe and coupling lasers. To avoid confusion with the main text, here we continue to refer to the lasers as pump and Stokes.

\begin{figure}
	\includegraphics[width=\columnwidth]{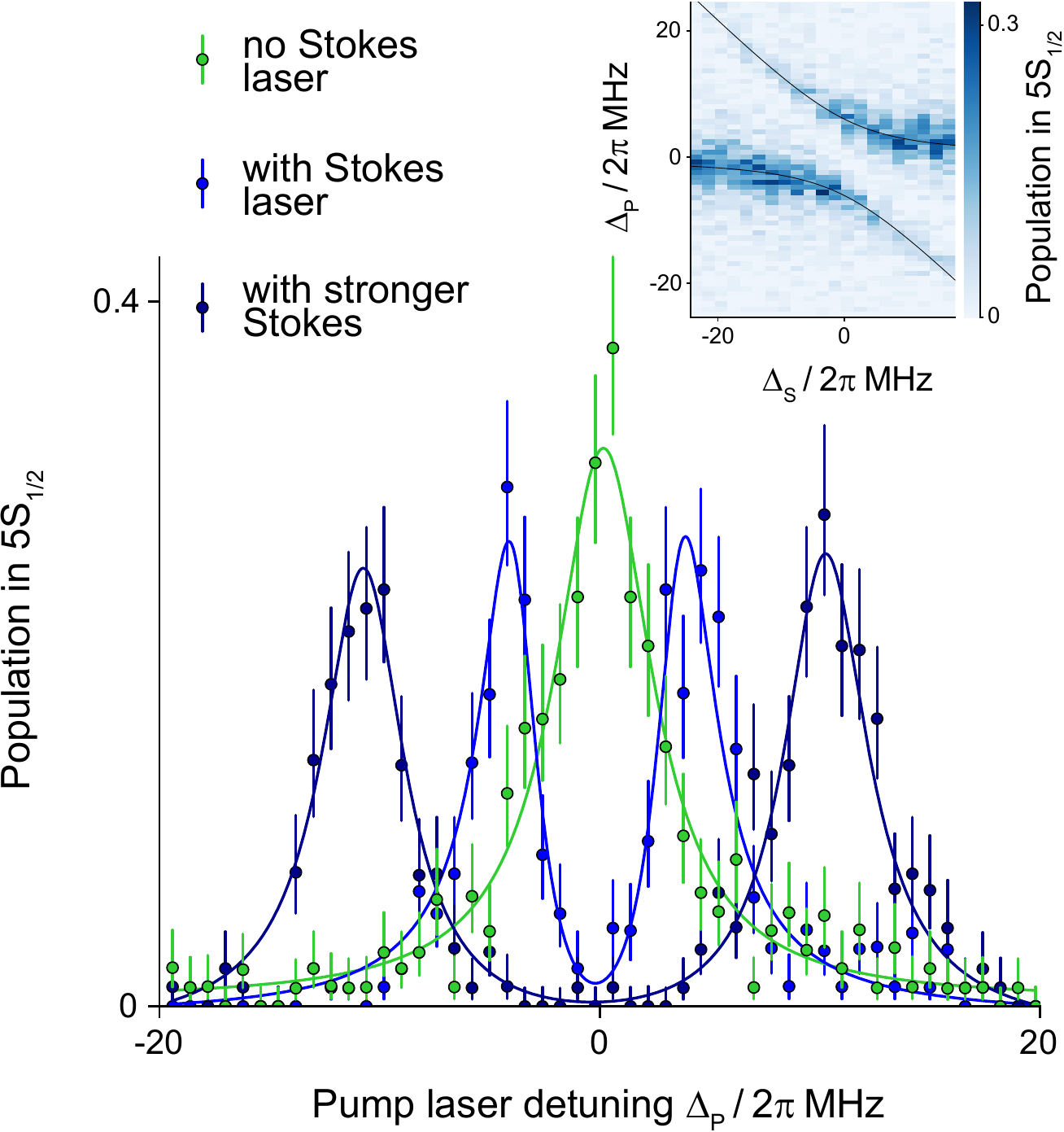}
	\caption{
		\textbf{Autler-Townes splitting is observed in the three-level system.}
		The ion is prepared in $|0\rangle$ and illuminated by the pump laser.
		In the absence of the Stokes laser field, the Lorentzian absorption profile of $|0\rangle\leftrightarrow |e\rangle$ is observed.
		When the resonant Stokes laser is switched on an Autler-Townes doublet emerges, with splitting $\Omega_{S}$.
		Error bars indicate projection noise (68\% CI).
		Inset:
		the $|e\rangle$ and $|r\rangle$ resonances exhibit an avoided crossing as a function of the Stokes laser detuning $\Delta_{S}$ and the pump laser detuning $\Delta_{P}$.
		Resonance positions follow the ac Stark formula $\Delta_{P}=\tfrac{1}{2}(-\Delta_{S}\pm \sqrt{\Delta_{S}^2+\Omega_{S}^2})$ with $\Omega_{S}=2 \pi \times 12.1\,\mathrm{MHz}$ shown in black lines.
		\label{Fig_supp}}
\end{figure}

The Lorentzian absorption profile of $|0\rangle \leftrightarrow |e\rangle$ is observed by scanning the frequency of the pump laser across the resonance in the absence of the Stokes laser (Fig.~\ref{Fig_supp}).
When the strong Stokes laser is turned on the absorption profile shows an Autler-Townes doublet.
The Stokes laser couples the $|e\rangle$ and $|r\rangle$ states, such that they are no longer energy eigenstates of the system.
The new energy eigenstates, dressed by the Stokes laser, are found by diagonalizing the coupling Hamiltonian $\mathsf{H}$ in the rotating wave approximation
\begin{equation*}
\mathsf{H}=\frac{\hbar}{2}
\begin{pmatrix}
    0          & \Omega_{P}            & 0                       \\
    \Omega_{P} & 2\Delta_{P}           & \Omega_{S} e^{i\phi}    \\
    0          & \Omega_{S} e^{-i\phi} & 2\Delta_{P}+2\Delta_{S}
\end{pmatrix}
\end{equation*}
in the basis $\{|0\rangle, |e\rangle, |r\rangle\}$.
When $\Delta_{S}=0$ and $\Omega_{S}\gg\Omega_{P}$ there appear two dressed eigenstates $|\phi^{\pm}\rangle=\frac{1}{\sqrt{2}}(|e\rangle \pm |r\rangle)$ which are shifted in energy from the bare states by $\pm \tfrac{1}{2}\hbar\Omega_{S}$.
Each peak in the Autler-Townes doublet corresponds to excitation of one of the dressed states.
Thus the splitting of the doublet increases with the strength of the Stokes laser.
This marks the first reported coherent effect observed with Rydberg ions.

The UV excitation time is much larger than the lifetimes of $|e\rangle$ and the dressed states $|e\rangle \pm |r\rangle$, thus the experiment may be modelled as optical pumping from $|0\rangle$ to $5S_{1/2}$ by incoherent excitation to $|e\rangle$ or to a superposition of $|e\rangle$ and $|r\rangle$, followed by decay to $5S_{1/2}$.
The data is fit by $1-e^{\mathcal{R}T_{ex}}$, with $\mathcal{R}$ the rate of absorption.
For the green curve $\mathcal{R}$ is given by a Lorenztian function, which is the natural radiative line shape of $|e\rangle$.
For the blue curves showing Autler-Townes splitting $\mathcal{R}$ is obtained by solving the Lindblad master equation for the four-level system \{$|0\rangle, |e\rangle, |r\rangle, |5S_{1/2}\rangle$\} by adiabatically eliminating $|e\rangle$ and $|r\rangle$. The absorption rate depends on $\Omega_{S}, \Omega_{P}, \Delta_{S}, \Delta_{P}, \Gamma_{e}$ and $\tau_{42S}$.

By scanning the frequencies of both UV lasers the avoided crossing of $|e\rangle$ and $|r\rangle$ is mapped out (Fig.~\ref{Fig_supp} inset).
The smooth transition from bare states to dressed states is displayed.
When $|\Delta_{S}| \gg \Omega_{S}$ the two resonances correspond to excitation of $|e\rangle$ at $\Delta_{P}=0$ and $|r\rangle$ when $\Delta_{P}=-\Delta_{S}$.
When $|\Delta_{S}| \not\gg \Omega_{S}$ the resonances correspond to excitation of dressed states.

We follow the sign convention $\Delta_{P},\Delta_{S}>0$ for UV lasers red-detuned from resonance.

\renewcommand{\bibnumfmt}[1]{[S#1]}
\end{document}